%% file: MorphPaper.tex
\documentclass[11pt]{myArticle}       

\pdfoutput=1

\usepackage{amsmath,amssymb,amsfonts} 
\usepackage{graphicx}
\usepackage{epsfig}
\usepackage{color}                    
\usepackage{hyperref}                 
\usepackage{myFancyhdr}               
\usepackage{makeidx}                  
\usepackage{helvet}                   
\usepackage{setspace}                 
\usepackage{fancybox}                 
\usepackage{float}                    
\usepackage{xspace}                   
\usepackage{wasysym}                  
\usepackage[numbers,sort&compress]{natbib}
\usepackage{hypernat}
\usepackage{wrapfig}
\usepackage{subcaption}
\usepackage{units}
\usepackage{lineno}
\usepackage{listings}

\definecolor{mygray}{rgb}{0.3,0.32,0.35}
\definecolor{darkblue1}{rgb}{0,0,.2}
\definecolor{darkblue}{rgb}{0,0,.3}
\definecolor{darkred}{rgb}{0.5,0,0}
\pagecolor{white} 
\color{black}     
\hypersetup{breaklinks=true,
            colorlinks=true,
            linkcolor=darkblue1,
            menucolor=darkblue1,
            urlcolor=darkblue1,
            citecolor=darkblue1,
            pdftitle={Non-linear moment morphing of multi-dimensional histograms},
            pdfauthor={RooFit Morph Group},
            pdfsubject={Non-linear moment morphing of multi-dimensional histograms},
            pdfkeywords={},
            pdfproducer={RooFit Morph Group}
}
\bibstyle{plain}
\newcommand\allFontSize{\small}

\newenvironment{myquote}
               {\list{}{\leftmargin0cm}%
                \item\relax}
               {\endlist}
\usepackage{caption}

\newcommand\detailsSize{\allFontSize}
\newenvironment{details}%
{\begin{myquote}\vspace{-0.2cm}\detailsSize}{\end{myquote}\vspace{-0.2cm}}

\newcommand{\CC}{C\nolinebreak\hspace{-.05em}\raisebox{.4ex}{\tiny\bf +}\nolinebreak\hspace{-.10em}\raisebox{.4ex}{\tiny\bf +}}

\makeindex
\makeatletter
\newcommand*{\bdiv}{%
  \nonscript\mskip-\medmuskip\mkern5mu%
  \mathbin{\operator@font div}\penalty900\mkern5mu%
  \nonscript\mskip-\medmuskip
}
\makeatother

\begin{document}
%
%
\pagenumbering{arabic}
\input Title
\newpage
%
%

\input Maintext

\addcontentsline{toc}{section}{References}
\bibliography{References}{}

\end{document}

%% file: Title.tex
{\small
\color{mygray}
\begin{flushright}
{\sf\em CERN-OPEN-2014-050} \\
{\sf\em \today} \\
\end{flushright}
}
\def\UrlFont{\rm}

\vspace{1.3cm}

{\sf\LARGE\bfseries
Interpolation between multi-dimensional histograms using a new non-linear moment morphing method}

\vspace{1.0cm}

{\Large
  M.~Baak$^{a}$, S.~Gadatsch$^{b}$, R.~Harrington$^{c}$, W.~Verkerke$^{b}$
}

\vspace{0.5cm}

{\normalsize
  $^{a}$CERN, Geneva, Switzerland \\
  $^{b}$Nikhef, Amsterdam, The Netherlands \\
  $^{c}$University of Edinburgh, Edinburgh, Scotland

\vspace{1.0cm}

\begin{details} {\sf\bfseries Abstract}

A prescription is presented for the interpolation between multi-dimensional distribution
templates based on one or multiple model parameters.
The technique uses a linear combination of templates, each created using fixed values of the model's parameters and transformed
according to a specific procedure, to model a non-linear dependency on model parameters and the dependency between them.
By construction the technique scales well with the number of input templates used, which is a useful
feature in modern day particle physics, where a large number of templates is often required to model
the impact of systematic uncertainties.
\end{details}

\thispagestyle{empty}

%% file: Maintext.tex
\input Intro

\section{Construction of the morphing p.d.f.} \label{sec:construction}

This section details the construction of the moment morphing probability density function (p.d.f.).
The method proposed here is based on the linear combination of input templates.
The dependency on the morphing parameter(s) can be non-linear, and is captured
in multiplicative coefficients and a transformation of the template observables.
Interpolation using a single morphing parameter is described in Section~\ref{sec:oneparameter}.
Section~\ref{sec:nparameters} describes interpolation using multiple morphing parameters and
shows that dependencies between morphing parameters can be readily modeled.
Other choices of basis functions
for the construction of the morphing p.d.f. are considered in Section~\ref{sec:basis}.

\subsection{Interpolation with a single morphing parameter}
\label{sec:oneparameter}

Consider an arbitrary p.d.f. $f(\mathbf{x}|m)$, where $f$ depends on the single morphing parameter $m$ and describes the observables $\mathbf{x}$.
The true dependency on $m$ is not known or difficult to obtain.
Instead, the p.d.f. $f$ has been sampled at $n$ different values of $m$, with each $f(\mathbf{x}|m_i)$ representing a known
input template shape for a single value of the morphing parameter, labeled $m_i$\,.
In the following the goal is to construct a parametric approximation of $f(\mathbf{x}|m)$ for arbitrary $m$, which is continuous and smooth in the model parameter,
as required for example by the statistical tests used in particle physics alluded to in Section~\ref{sec:intro}.
There are two steps to this.

First, given the sampling points, $f(\mathbf{x}|m)$ can be expanded in a Taylor series up to order $n-1$ around reference value $m_0$,
\begin{equation}\label{firsteq}
  f(\mathbf{x}|m) \approx \sum_{j=0}^{n-1} \frac{{\mathrm d}^{(j)} f(\mathbf{x}|m_0)}{{\mathrm d} m^{(j)}} \frac{(m - m_0)^j}{j!}
  = \sum_{j=0}^{n-1} f^{\prime}_j(\mathbf{x}|m_0) (m - m_0)^j\,,
\end{equation}
where the second equality defines $f^\prime(\mathbf{x}|m)$.
For the $n$ given values of $m$ follows the vector equation:
\begin{equation}\label{veceqn}
  f(\mathbf{x}|m_i) \approx \sum_{j=0}^{n-1} (m_i - m_0)^j  f^{\prime}_j(\mathbf{x}|m_0)
  = \sum_{j=0}^{n-1} M_{ij} f^{\prime}_j(\mathbf{x}|m_0)\,,
\end{equation}
where $M_{ij} = (m_i - m_0)^j$ defines a $n \times n$ transformation matrix.
Inverting Eqn.~\ref{veceqn} gives
\begin{equation}
  f^{\prime}_j(\mathbf{x}|m_0) = \sum_{i=0}^{n-1} \left(M^{-1}\right)_{ji} f(\mathbf{x}|m_i)
\end{equation}
which allows to determine the $n$ values $f^{\prime}_j(\mathbf{x}|m_0)$. Substituting this in Eqn.~\ref{firsteq}, $f(\mathbf{x}|m)$ reads
\begin{equation}
  f(\mathbf{x}|m) \approx \sum_{i,j=0}^{n-1} (m - m_0)^j \left(M^{-1}\right)_{ji} f(\mathbf{x}|m_i)\,,
\end{equation}
which can be used to predict the template shape at any new value of the morphing parameter given by $m^\prime$,
\begin{equation}\label{newpoint}
  f_\text{pred}(\mathbf{x}|m^\prime) = \sum_{i=0}^{n-1} c_i(m^\prime) f(\mathbf{x}|m_i)\,,
\end{equation}
which is a linear combination of the input templates $f(\mathbf{x}|m_{i})$, each multiplied by a coefficient $c_i(m^\prime)$,
\begin{equation} \label{eq:coefficient}
  c_i(m^\prime) = \sum_{j=0}^{n-1} (m^\prime - m_0)^j \left(M^{-1}\right)_{ji}\,,
\end{equation}
which themselves are non-linear and depend only on the distance to the reference points.
This approach of weighting the input templates is also known as vertical morphing.
Note that the coefficients $c_i$ are independent of the derivatives of $\mathbf{f}$ with respect to morphing parameters or to the observable set $\mathbf{x}$, making their computation easy.

The coefficient for a point included in the set of input templates is one, \emph{i.e.}
\begin{equation}
c_i(m_j) = \delta_{ij}\,,
\end{equation}
and by construction the sum of all coefficients $c_i$ equals one:
\begin{equation}
\sum_i c_i(m) = 1.
\end{equation}
This turns out to be a useful normalization, as will be seen below.

To illustrate, one can consider a morphed p.d.f. using only input templates at two values of the morphing parameter, $m_\text{min}$ and $m_\text{max}$.
The coefficients $c_i(m)$ become linear in $m$ and reduce to the simple fractions:
\begin{eqnarray} \label{eq:coefficientlinear}
        c_{i_\text{min}} &=& 1 - m_\text{frac}\\
        c_{i_\text{max}} &=& m_\text{frac}\,,
\end{eqnarray}
where $m_\text{frac}=(m-m_\text{min})/(m_\text{max}-m_\text{min})$, $c_{i_\text{min}}$ and $c_{i_\text{max}}$ sum up to one, and all other coefficients are zero.

Second, it may be that the sampled input p.d.f.s $f_i$ describe distributions in $\mathbf{x}$ that vary strongly
as a function of $m$ in shape and location.
This is equivalent to the first and second moments (\emph{i.e.} the means and variances) of the input
distributions having a dependence on the morphing parameter $m$.

Since the input p.d.f.s in Eqn.~\ref{newpoint} are summed linearly, it is imperative to translate all input distributions $f_i(\mathbf{x})$ in the sum before
combining in the morphed p.d.f. such that their locations match up.
The process of translating the input observables (but not scaling; see below) is also called horizontal morphing.
In addition it is necessary to take in account the change in the width of the input distributions as a function of the morphing parameter.

To achieve this,
the mean $\mu_{ij}$ and width $\sigma_{ij}$ of each input distribution $i$ and observable $x_j$ are shifted to the common values of $\mu'_{j}(m)$ and $\sigma'_{j}(m)$.
These are obtained by
multiplying the underlying means and widths with the coefficients $c_i(m)$ of Eqn.~\ref{eq:coefficient} 
\begin{align}\label{newmean}
 \mu'_{j}(m) &= \sum_i c_i(m) \cdot \mu_{ij}\\
 \sigma'_{j}(m) &= \sum_i c_i(m) \cdot \sigma_{ij}
\end{align}
In order to shift the input p.d.f.s a linear transformation of each observable is applied.
For each p.d.f. $i$ and observable $j$ define
\begin{equation}
x'_{ij} = a_{ij} x_{j} + b_{ij}\,,
\end{equation}
with slope
\begin{equation}
a_{ij}=\frac{\sigma_{ij}}{\sigma'_j}
\end{equation}
and offset
\begin{equation}
 \quad b_{ij}=\mu_{ij}-\mu'_j a_{ij}\,.
\end{equation}
to replace the original observables $x_{j}$ in the input p.d.f.s
\begin{equation}
 f(\mathbf{x}|m_i) \rightarrow f(\mathbf{x'}|m_i)\,.
\end{equation}

Since only a linear transformation is applied to each observable,
\emph{i.e.} only the first two moments of the input p.d.f.s are modified,
the normalization of the scaled input template is analytically related to the normalization of the original
template as
\begin{equation} \label{eq:selfnormalization}
 \int_{-\infty}^{+\infty} f(\mathbf{x'}|m_i) d\mathbf{x} = \frac{1}{\prod_j a_j(m)}\,\int_{-\infty}^{+\infty} f_i(\mathbf{x}|m_i)d\mathbf{x}\,,
\end{equation}
with the slope $a_j = \sigma_j / \sigma^\prime_j$, where $j$ refers to the observable $x_j$.
The construction of the complete morphed p.d.f. as the sum of the transformed input p.d.f.s then gives
\begin{equation} \label{eq:morphpdf}
p(\mathbf{x}|m')=\sum_i c_i(m')  f(\mathbf{x'},m_i) \prod_j a_j(m') \,.
\end{equation}
As the coefficients $c_i$ add up to $1$ by construction,
the morphed p.d.f. of Eqn.~\ref{eq:morphpdf} is unit-normalized as well for normalized input templates.

This leads to an important computational advantage and novelty: for models where the input templates are constant,
such as histogram-based templates, no normalization integrals need to be recalculated during the
minimization of the likelihood function, which is often a bottle-neck when using morphed p.d.f.s.
As a result, the number of input templates is generally increasable without significant performance loss.

Note that the self-normalization of Eqn.~\ref{eq:selfnormalization} remains
valid when applying a rotation to the set of (multiple) observables,
which would introduce covariance moments to the modified input p.d.f.s.
Though technically possible, such rotations are avoided here as they obscure
the physical interpretation of the observable set.
A consequence of this on the accuracy of the morphed p.d.f. to model changing
correlations between observables is discussed in Section~\ref{sec:accumult}.

\begin{figure}[htp]
  \centering
  \begin{minipage}[b]{.49\linewidth}
    \centering
    \includegraphics[width=\textwidth]{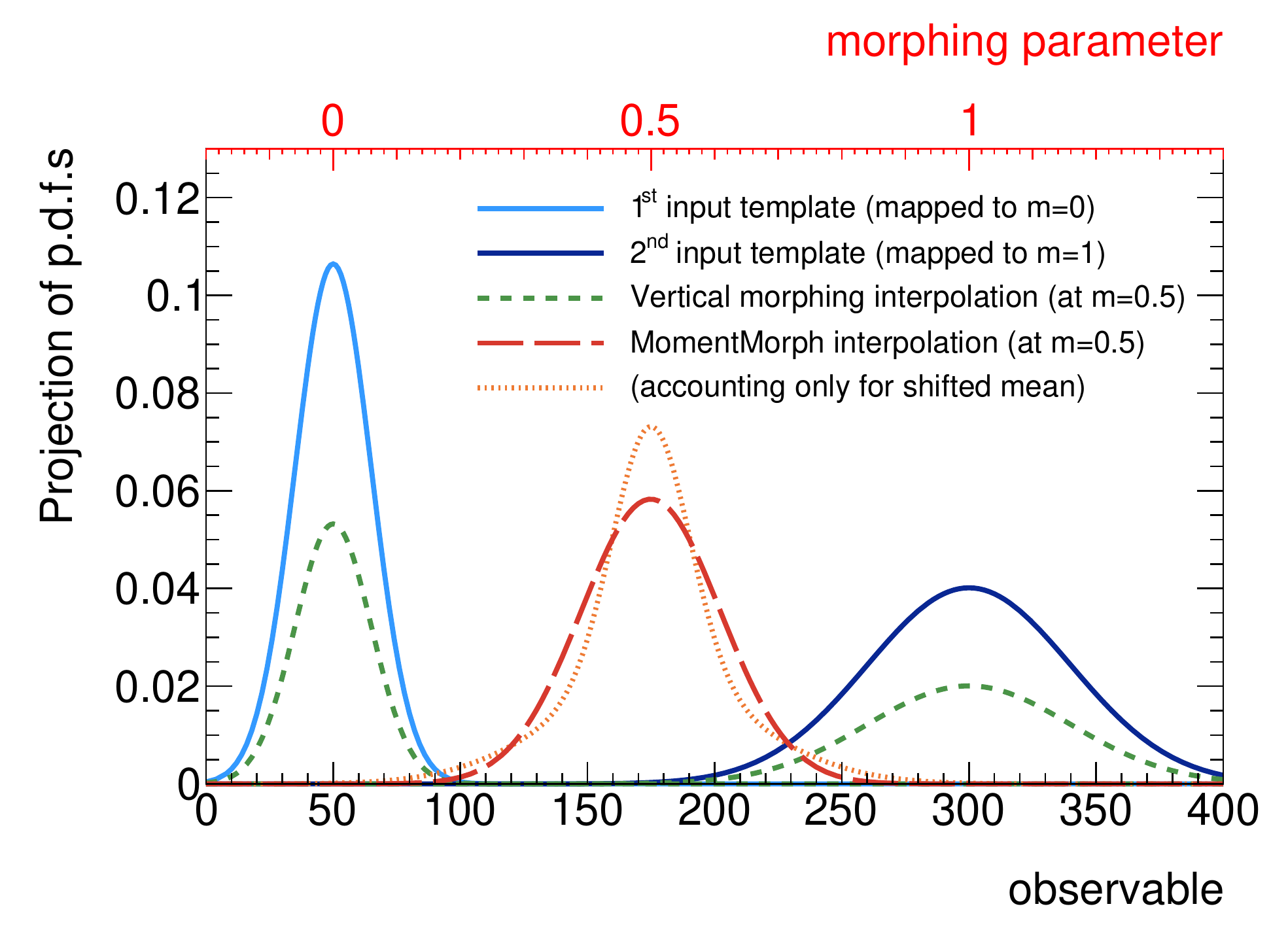}
    \subcaption{}\label{fig:construction}
  \end{minipage}%
  \begin{minipage}[b]{.49\linewidth}
    \centering
    \includegraphics[width=\textwidth]{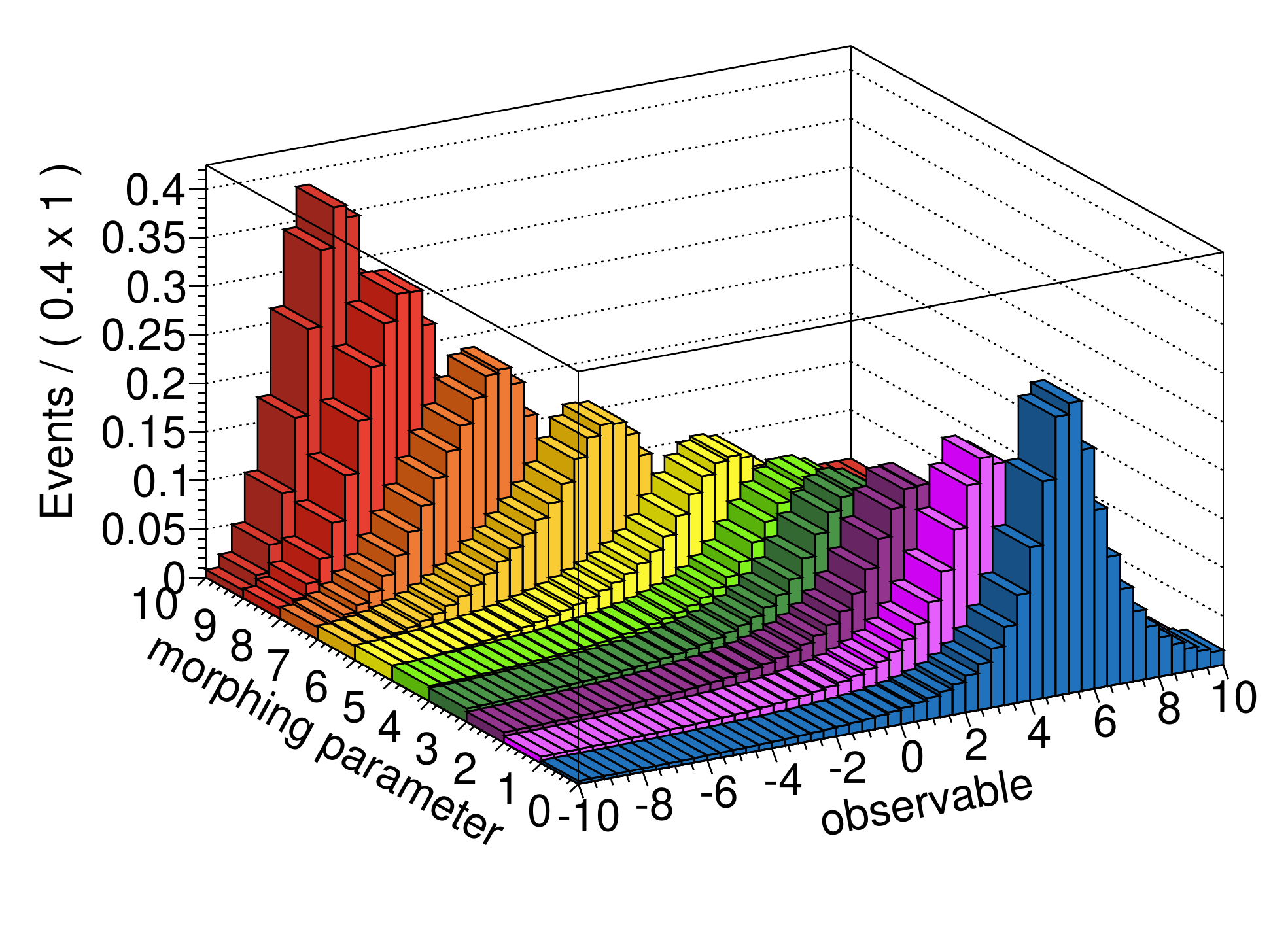}
    \subcaption{}\label{fig:example1}
  \end{minipage}
  \caption{Two examples of moment morphing. (\subref{fig:construction})~Construction of the morphed p.d.f. The interpolation is done between two normal distributions, shown as solid lines, corresponding to values $0$ and $1$ of the morphing parameter. After vertical morphing, the mean of the templates is shifted to the common value and their widths are adjusted accordingly. The dashed p.d.f. shows the morphed p.d.f., which is a linear combination of the modified inputs. (\subref{fig:example1})~Non-linear morphing of a Cauchy distribution ($m = 0$) via a Crystal Ball line shape ($m = 5$) into a normal distribution ($m = 10$).}
\end{figure}

The processes of vertical and horizontal morphing (\emph{i.e.} summing and translating) and of scaling the input morphed p.d.f.s are illustrated in Fig.~\ref{fig:construction},
which morphs between two normal distributions.
The technique proposed also accurately models the evolution of rapidly changing distributions as illustrated in Fig.~\ref{fig:example1}.
In the sample, the application of moment morphing is used to describe the non-linear transition of a Cauchy distribution via a Crystal Ball line shape into a normal distribution. The parameters of the used p.d.f.s are chosen such that the positions of their means as well as their shapes vary substantially as a function of the morphing parameter $\alpha$, in particular in the tails of the distributions which change dynamically along the morphing path.

\begin{figure}[htp]
  \centering
  \begin{minipage}[b]{.6\linewidth}
    \centering
    \includegraphics[width=\textwidth]{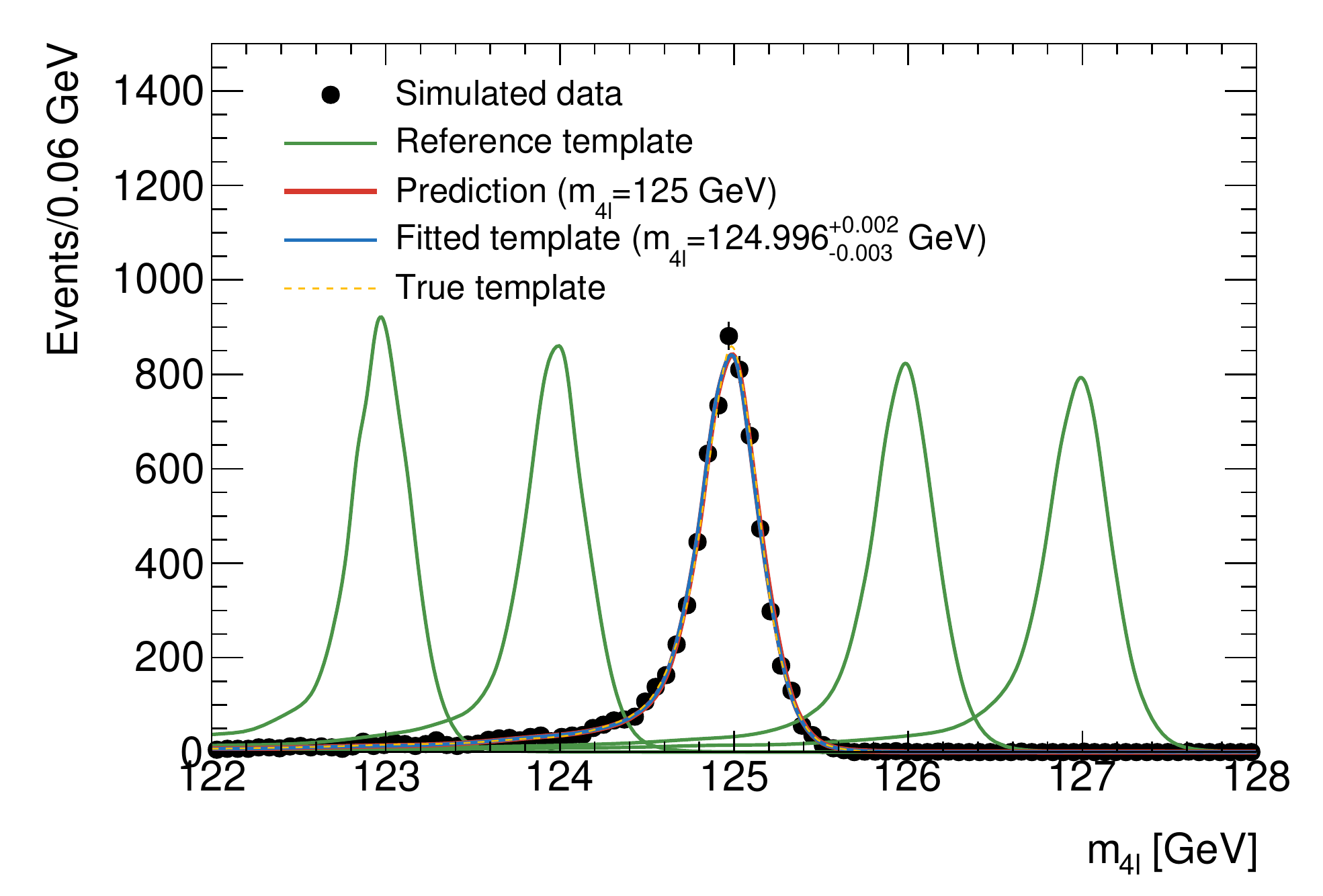}
  \end{minipage}
  \caption{Example of p.d.f. interpolation. The reconstructed invariant mass distribution of for a $\unit[125]{GeV}$ Standard Model Higgs boson decaying to four leptons (solid, red) has been predicted by non-linearly interpolating between the reference distributions for surrounding mass hypotheses (solid, green), as described in the text. The prediction is compared to the true $\unit[125]{GeV}$ template (dashed, yellow) derived from the simulated data for this hypothesis and a fit of the morphed p.d.f. to this dataset (solid, blue). The predicted, true, and fitted curves match almost perfectly.}
  \label{fig:prediction}
\end{figure}

Fig.~\ref{fig:prediction} shows an application of the technique described in this article to a complex physics and detector simulation.
The reconstructed invariant mass distribution for a Standard Model Higgs boson with mass $\unit[125]{GeV}$ decaying to four leptons is described by non-linearly interpolating between a series of templates corresponding to simulation response estimates for four assumed Higgs boson masses, at $123$, $124$, $126$ and $\unit[127]{GeV}$.
As reference templates kernel estimation p.d.f.s modeling events simulated with MadGraph~\cite{Alwall:2014hca} and an ATLAS-type PGS~\cite{PGSWeb}
simulation are used.
Despite rapidly evolving features, the template predicted by the morphing technique reproduces the true template at $\unit[125]{GeV}$.
The resulting morphed p.d.f. is parametrized in terms of the ``true'' Higgs boson mass, as opposed to the reconstructed invariant mass, 
and a fit of the morphed p.d.f. to a dataset with an assumed true Higgs boson of $\unit[125]{GeV}$ directly and accurately measures that 
true Higgs boson mass.

\subsection{Interpolation with multiple morphing parameters}
\label{sec:nparameters}

The non-linear implementation of multiple morphing parameters is analogous to the single parameter case of Section~\ref{sec:oneparameter}.
We illustrate here the expansion to two parameters,
including correlated effects in the template distribution caused by changing of two or more morphing parameters simultaneously,
before generalizing the approach to an arbitrary number of parameters.

The general Taylor series expansion of $f(\mathbf{x}|m_1,m_2)$, depending on parameters $m_1$ and $m_2$, around $(m_{1_0},m_{2_0})$ reads:
\begin{equation}
 \begin{split}
  f(\mathbf{x}|m_{1},m_{2}) & = f(\mathbf{x}|m_{1_0},m_{2_0})\\
  & + \frac{1}{1!} \left[(\Delta m_{1}) \frac{\partial}{\partial m_1} f(\mathbf{x}|m_{1_0},m_{2_0}) + (\Delta m_{2}) \frac{\partial}{\partial m_2} f(\mathbf{x}|m_{1_0},m_{2_0})\right]\\
  & + \frac{1}{2!} \left[ (\Delta m_{1})^2 \frac{\partial^2}{\partial m_1^2} f(\mathbf{x}|m_{1_0},m_{2_0}) + 2(\Delta m_{1})(\Delta m_{2}) \frac{\partial^2}{\partial m_1 \partial m_2} f(\mathbf{x}|m_{1_0},m_{2_0})\right.\\
  & +\left.(\Delta m_{2})^2 \frac{\partial^2}{\partial m_2^2} f(\mathbf{x}|m_{1_0},m_{2_0}) \right] + \ldots\,,
 \end{split}
\end{equation}
which, for a $2\times 2$ square grid with reference points surrounding $(m_1,m_2)$, is approximated by:
\begin{equation}
 \begin{split}
  f(\mathbf{x}|m_{1},m_{2}) & = f(\mathbf{x}|m_{1_0},m_{2_0}) \\
  & + (\Delta m_{1}) \frac{\partial}{\partial m_1} f(\mathbf{x}|m_{1_0},m_{2_0}) + (\Delta m_{2}) \frac{\partial}{\partial m_2} f(\mathbf{x}|m_{1_0},m_{2_0}) \\
  & + (\Delta m_{1})(\Delta m_{2}) \frac{\partial^2}{\partial m_1 \partial m_2} f(\mathbf{x}|m_{1_0},m_{2_0}) \,.
 \end{split}
\end{equation}
Here the middle two terms are the linear expansions along $m_1$ and $m_2$,
and the last term represents the expansion along both $m_1$ and $m_2$ simultaneously.
The addition of the fourth ``corner'' point allows one to model functions that cannot be factorized
into functions depending only on $m_1$ or $m_2$.

Repeating the Taylor series expansion for a grid of $k \times l$ reference points, and again writing the truncated series as a vectorial equation,
leads  to Eqn.~\ref{veceqn} with additional mixed terms in the transformation matrix $M$ and the derivatives $f^\prime_j$.
The transformation matrix $M$ now reads
\begin{equation}
  M = \begin{pmatrix}
    1 & 0  & \cdots & 0 & 0 & \cdots & 0 & \cdots & 0 \\
    1 & (\Delta m_{2_{10}})  & \cdots & (\Delta m_{2_{10}})^k & (\Delta m_{1_{10}}) & \cdots & (\Delta m_{1_{10}})(\Delta m_{2_{10}})^k & \cdots & (\Delta m_{1_{10}})^l(\Delta m_{2_{10}})^k \\
    \vdots & \vdots & \cdots & \vdots & \vdots & \cdots & \vdots & \cdots & \vdots \\
    1 & (\Delta m_{2_{n0}})  & \cdots & (\Delta m_{2_{n0}})^k & (\Delta m_{1_{n0}}) & \cdots & (\Delta m_{1_{n0}})(\Delta m_{2_{n0}})^k & \cdots & (\Delta m_{1_{n0}})^l(\Delta m_{2_{n0}})^k
  \end{pmatrix}\,,
\end{equation}
where $\Delta m_{i_{n0}} = m_{i_n} - m_{i_0}$ is a short notation for the distance between reference point $n$ and reference point $0$ in the $i\text{th}$ dimension of the parameter space.
The distance vector $\mathbf{\Delta m}$ is in multiple dimensions defined as
\begin{equation}
  \mathbf{\Delta m} = \begin{pmatrix}
    1 & (\Delta m_{2_{l0}}) & \cdots & (\Delta m_{2_{l0}})^k & (\Delta m_{1_{l0}}) & \cdots & (\Delta m_{1_{l0}}) (\Delta m_{2_{l0}})^k  & \cdots & (\Delta m_{2_{l0}})^l(\Delta m_{2_{l0}})^k
  \end{pmatrix}^T \,.
\end{equation}
Following Eqn.~\ref{newpoint}, the coefficients $c_i$ for a new point $(m^\prime_{1}, m^\prime_{2}) = (m_{1_q}, m_{2_q})$ are now given by
\begin{equation}
 c_i(m^\prime_{1},m^\prime_{2}) = \sum_{j=0}^{(k\times l)-1} \left(M^{-1}\right)_{ji} \cdot (\Delta m)_j\,.
\end{equation}
The construction of the morphed p.d.f. $p(\mathbf{x}|m_1,m_2)$ using these coefficients is as in Eqn.~\ref{eq:morphpdf}.

The extension to an arbitrary number of morphing parameters, $n$, is a matter of using an $n$-dimensional
grid of input p.d.f.s, with $k\times l\times \ldots$ grid points,
and consistently expanding the transformation matrix $M$ with the additional higher order terms.

Returning to the $2\times 2$ square grid of input p.d.f.s with reference values surrounding $(m_1,m_2)$,
the coefficients in $(m_1,m_2)$ reduce to
\begin{eqnarray}
        c_{00}(m_1,m_2) &=& (1 - m_\text{1,frac})\cdot (1 - m_\text{2,frac}) \\
        c_{10}(m_1,m_2) &=& m_\text{1,frac}\cdot (1 - m_\text{2,frac}) \\
        c_{01}(m_1,m_2) &=& (1 - m_\text{1,frac})\cdot m_\text{2,frac} \\
        c_{11}(m_1,m_2) &=& m_\text{1,frac}\cdot  m_\text{2,frac} \,,
\end{eqnarray}
with $m_\text{1,frac}=(m_{1}\!-\!m_{1_{00}})/(m_{1_{10}}\!-\!m_{1_{00}})$ and $m_\text{1,frac}=(m_{2}\!-\!m_{2_{00}})/(m_{2_{01}}\!-\!m_{2_{00}})$.
Note that the coefficients are all positive when staying within the boundaries of the grid, and add up to $1$.
Towards the corner point $(m_{1_{11}},m_{2_{11}})$, the nominal coefficient $c_{00}$
and linear-expansion coefficients $c_{10}$ and $c_{01}$ are turned off, and the
quadratic term $m_\text{1,frac}\cdot m_\text{2,frac}$ in $c_{11}$ is turned on in full,
describing the change caused by changing $m_1$ and $m_2$ simultaneously.

\begin{figure}[!t]
  \centering
  \begin{minipage}[b]{.6\linewidth}
    \centering
    \includegraphics[width=\textwidth]{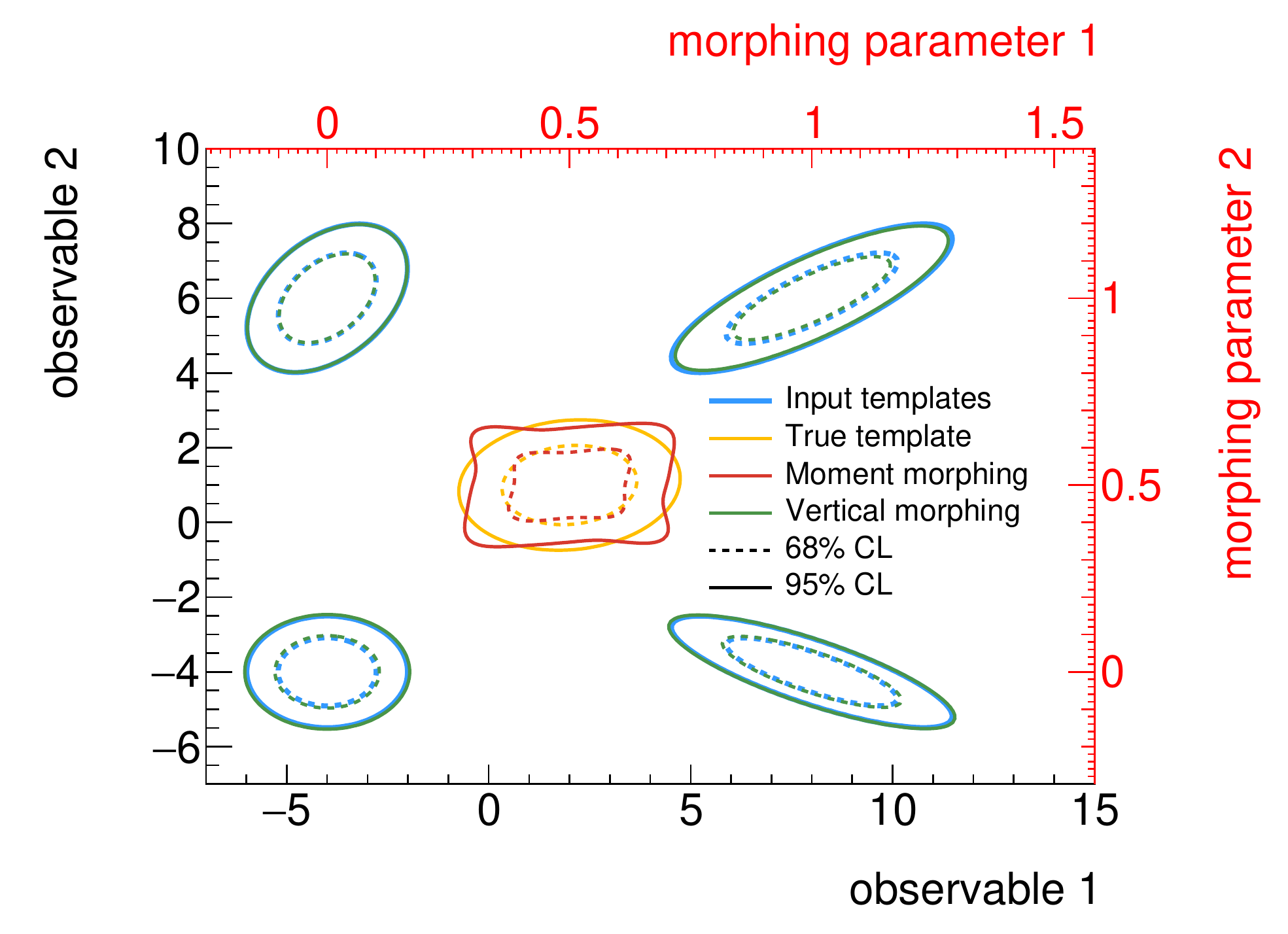}
  \end{minipage}
  \caption{Example of multi-dimensional moment morphing. Linear interpolation between multivariate normal distributions in a multi-dimensional parameter space.
    For illustration purposes only, the plot has a 1-to-1 mapping between the corners of the morphing parameter and observable space.
    The example is constructed such that mean and covariance of the multivariate normal distributions follow an analytic description, in particular the variation of Pearson's correlation coefficient can be described bi-linearly. The corner contours (blue) of the grid represent the input templates to the morphing algorithm. Dashed and solid contours indicate one and two standard deviations from the mean respectively. The true template in the center of the grid, \emph{i.e.} at $(0.5, 0.5)$, is shown in yellow. The prediction of the morphed p.d.f. for the central point is shown in red. The green contours represent the prediction from vertical interpolation.
  }\label{fig:example2}
\end{figure}

Shown in Fig.~\ref{fig:example2} is an illustration of a multivariate normal distribution modeling the dependency between two observables.
The covariance matrix as well as the mean change as a function of two parameters $\alpha_1$ and $\alpha_2$.
In the given example the distributions are known for $2\times 2$ grid points fulfilling $\alpha_i = \lbrace 0, 1 \rbrace,\ i = 1,2$.
The information is used to linearly interpolate the p.d.f. to any desired point in the two-dimensional parameter space.
The yellow contours in the center of the grid, \emph{i.e.} at $(\alpha_1, \alpha_2) = (0.5, 0.5)$, represent the true template,
which is compared with the prediction by the moment morphed p.d.f. (red) and the prediction by vertical morphing only (green).
Similar to the example presented in Fig.~\ref{fig:construction}, vertical morphing only does not yield a central template with shifted mean.
The difference between the moment morphed template and the true template is induced by the change of the correlation between the two observables with varying morphing parameter values.
In case of constant correlation between the observables of the input templates, the prediction of the moment morphed p.d.f. is exact.
More discussion on this follows in Section~\ref{sec:accumult}.

\input basis

\input starmorph

\input accuracy

\input implementation

\input conclusion

\section*{Acknowledgments}
\label{sec:Acknowledgments}

This work was supported by CERN, Switzerland; FOM and NWO, the Netherlands; and STFC, United Kingdom.

%% file: Intro.tex
\section{Introduction}
\label{sec:intro}

In particle physics experiments, data analyses generally use shapes of kinematical distributions of reconstructed particles to interpret
the observed data. These shapes are simulated using Standard Model or other theoretical predictions, and are determined separately
for signal and background processes.
Simulations of known fundamental physics processes are run through a detailed detector
simulation, and are subsequently reconstructed with the same algorithms as the observed data.
These simulated samples may depend on one or multiple model parameters,
for example the simulated Higgs particle mass,
and a set of such samples may be required to scan over the various parameter values.
Since Monte Carlo simulation can be time-consuming,
there is often a need to interpolate between the limited number of
available Monte Carlo simulation templates.

In particular, the statistical tests widely used in particle physics,
\emph{e.g.} for the construction of confidence intervals on model parameters or the discovery of new phenomena,
rely strongly on continuous and smooth parametric models that describe the signal and background processes in the data\footnote{Typically, a statistical test
involves the maximization of a likelihood function, which has been built from both the parametric model and the observed data.
The maximization procedure relies on the derivatives of the likelihood with respect to the model's parameters.}.
These parametric models describe parameters of interest, such as a shifting mass parameter or
the rate of a signal process, and so-called nuisance parameters
that parametrize the impact of systematic uncertainties.
As such, the models are often constructed in terms of those parameters
by interpolating between simulated Monte Carlo templates, thereby ensuring continuity in those parameters.

Several algorithms exist that can be used to interpolate between Monte Carlo sample distributions~\cite{Read:1999kh,Cranmer:2012sba}.
Interpolation techniques have been used on multiple occasions in particle physics,
for example to predict kinematic distributions for intermediate values of a model parameter,
\emph{e.g.} the simulated Higgs boson, $W$ boson or top quark mass,
or to describe the impact of systematic uncertainties, which are often modeled as shape or rate
variations about a nominal template of a kinematic distribution.

This work describes a new morphing technique, {\it moment morphing},
which has the advantage over existing methods in that it is fast, numerically stable,
allows for both binned histogram and continuous templates,
has proper vertical as well as horizontal morphing (explained in Section~\ref{sec:construction}),
and is not restricted in the number of input templates, the number of model parameters or the number of input observables.
In particular, the latter feature allows the moment morphing technique to model the impact of
a non-factorizable response between different model parameters, where varying one model parameter
at a time is insufficient to capture the full response function.

The paper is organized as follows: Section~\ref{sec:construction} describes in detail how the moment morphing
function, used to interpolate between histograms, is constructed using one or more morphing parameters.
Section~\ref{sec:pdf_with_systematics} describes how the moment morphing technique can be used to properly take
into account systematic uncertainties in a high energy physics analysis, giving an example of a typical application.
A comparison in terms of accuracy of moment morphing with alternative morphing algorithms is provided in Section~\ref{sec:accurary}. %
Section~\ref{sec:implementation} describes the implementation of the moment morphing algorithm in publicly
available C++ code, including benchmarking of its performance.

%% file: basis.tex

\subsection{Other choices of basis functions}
\label{sec:basis}

Finally, having derived the morphed p.d.f. using a Taylor series expansion,
the dependency on a morphing parameter $m$ can be easily re-expressed in any orthonormal basis $\lbrace \Psi_0, \ldots, \Psi_{n-1} \rbrace = \lbrace \Psi_i \rbrace_{i = 0, \dots, n-1}$. The choice of the basis functions can simplify the studied problem and thus lead to a better approximation, even to an exact description.
For example, if the dependency on model parameters is expected to be periodic, it can be expressed in Fourier space as a linear combination of the eigenvectors $\Psi_i$:
\begin{equation}
    p(\mathbf{x}|m) = \sum_{i = 0}^{n-1} d_i(\mathbf{x}) \Psi_i(m)\, ,
\end{equation}
where the coefficients $d_i$ read
\begin{equation}
    d_i(\mathbf{x}) = \int p(\mathbf{x}|m) \Psi_i(m) dm\, .
\end{equation}
Thus, the dependency of the morphed p.d.f. on the observables enters through the coefficients $d_i(\mathbf{x})$ only, while the basis functions are independent of $\mathbf{x}$.

Note that the number of sampling points limits the number of utilizable basis functions $\Psi_i(m)$.
Analogously to Eqn.~\ref{newpoint}, $p$ can then be predicted at any new point $m^\prime$,
\begin{equation}
  p_\text{pred}(\mathbf{x}|m^\prime) = \sum_{i,j=0}^{n-1} \Psi_j(m^\prime) \left(M^{-1}\right)_{ji} p(\mathbf{x}|m_i)\,,
\end{equation}
with the modified $n \times n$ transformation matrix $M_{ij} = \Psi_j(m_i)$.

%% file: starmorph.tex
\section{A p.d.f. for modeling systematic uncertainties}
\label{sec:pdf_with_systematics}

In a typical analysis performed in a particle physics experiment,
the impact of systematic uncertainties is typically quantified by varying one-by-one the model parameters relating to detector modeling and physics models (\emph{e.g.} energy calibrations, factorization scales) and to record the template distribution with these modified settings.
The resulting pairs of alternate templates, corresponding to ``up'' and ``down'' variations of the uncertain model parameters,
must then be incorporated in the likelihood model of the physics analysis,
in the form of a model nuisance parameter that causes the template distribution to deform
as prescribed by the pair of ``alternate'' templates.

For the most widely-used statistical test at the LHC, the profile likelihood ratio,
it is required that the modeling of the systematic uncertainty in terms of a nuisance parameters
is done in a continuous way.
In particular, the maximization of the likelihood function, as needed (twice) in the profile likelihood ratio,
requires continuous and smooth parametric models to describe the signal and background processes present in the data\footnote{Specifically,
the likelihood for a physics measurement, $L(\mu,\boldsymbol\eta)$,
where $\mu$ is the physics parameter of interest and $\boldsymbol\eta$ are the
nuisance parameters that parametrize the impact
of systematic uncertainties on the signal and background predictions,
must be defined for all values of $\mu$ and $\boldsymbol\eta$.
The profile likelihood ratio is given by
\begin{equation}
\lambda(\mu) = \frac{L(\mu,\hat{\hat{\boldsymbol\eta}})}{L(\hat{\mu},\hat{\boldsymbol\eta})}\,,
\label{stat_lambda_eq}
\end{equation}
where $\hat{\mu}$ and $\hat{\boldsymbol\eta}$ represent the unconditional maximum likelihood estimates of $\mu$ and $\boldsymbol\eta$,
and $\hat{\hat{\boldsymbol\eta}}$ represents the conditional maximum likelihood estimate for the chosen value of $\mu$.
Therefore $\lambda$ is a function of $\mu$.
(Note that the data is omitted in the short hand notation of $L$.)
For a more detailed discussion on the profile likelihood ratio test statistic, see Ref.~\cite{Cowan:2010js}.}.

This section builds a parametrized p.d.f. describing a set of systematic variations about
the nominal prediction for a signal or background process,
using the morphing technique of Section~\ref{sec:construction}. 
Each systematic uncertainty $i$ is described with a nuisance parameter, $\eta_i$, 
that continuously morphs between the variation and nominal templates such that $\eta_i=\pm 1$ corresponds to the
$\pm 1\sigma$ ($1\sigma$ = 1 standard deviation) variations, and
$\eta_i=0$ corresponds to the nominal template\footnote{
A Gaussian constraint is applied separately for each systematic uncertainty to account for
uncertainty in the external measurement.
This constraint preferably centers each systematic variation around the nominal prediction,
with a reduced likelihood for potential shifts;
however, a combined fit to the observed data may of course prefer shifts in the nuisance parameters.}.
Additional variation templates may be added for different values of $\eta_i$.

The response of the likelihood function to changes in the nuisance parameters is here assumed to be factorized,
\emph{i.e.} it is assumed that the effect of a simultaneous changes in two or more nuisance parameters can be
described as a superposition of the effects of changing each nuisance parameter individually.
(Unfactorizable uncertainties are not discussed here; their treatment is handled following the recipe of Section~\ref{sec:nparameters}.)
Where $n$ unfactorizable uncertainties would require a full $n$-dimensional grid of input templates,
this assumption reduces the number of required inputs to a set of $n$ one-dimensional variations,
with only the nominal template in common -- a ``star'' shape in the nuisance parameter space.

The construction of the morphed p.d.f. as the sum of the input templates becomes
\begin{equation} \label{eq:starmorphpdf}
 p_{\rm pred}(\mathbf{x}|\mathbf{\eta}) =  (1-\sum_{i=1}\sum_{j=\pm1,\pm m} c_{ij}(\eta_i)) \cdot p(\mathbf{x},0) +  \sum_{i=1}\sum_{j=\pm1,\pm m} c_{ij}(\eta_i) \cdot p_{ij}(\mathbf{x}|\eta_i=j) \,,
\end{equation}
where the double-sum runs over the implemented systematic uncertainties and their available $\pm1\,\ldots\, \pm m\,\sigma$ variations.
The morphed p.d.f. is self-normalized, as by construction all coefficients add up to one.

In the publicly available implementation, detailed in Section~\ref{sec:classes}, the coefficients $c_{ij}(\eta_i)$ are linear, and depend only
the two closest input points surrounding $\eta_i$, as in Eqn.~\ref{eq:coefficientlinear}.
Also, the linear transformation of the observables $\mathbf{x}\to\mathbf{x'}$,
responsible for scaling and horizontal morphing, can be turned on or off. (By default it is off.)
Henceforth this p.d.f. is called the star-morphed p.d.f.

This type of star-morphed p.d.f. has been used in the analysis of Higgs decay to 4 leptons ($H \rightarrow ZZ^{*} \rightarrow 4l$) by
the ATLAS experiment~\cite{Aad:2014eva}
to describe the various background components contributing the $ZZ$ mass spectrum,
each including the variations of all relevant systematic uncertainties.

\begin{figure}[!htp]
  \centering
  \begin{minipage}[b]{.60\linewidth}
    \centering
    \includegraphics[width=\textwidth]{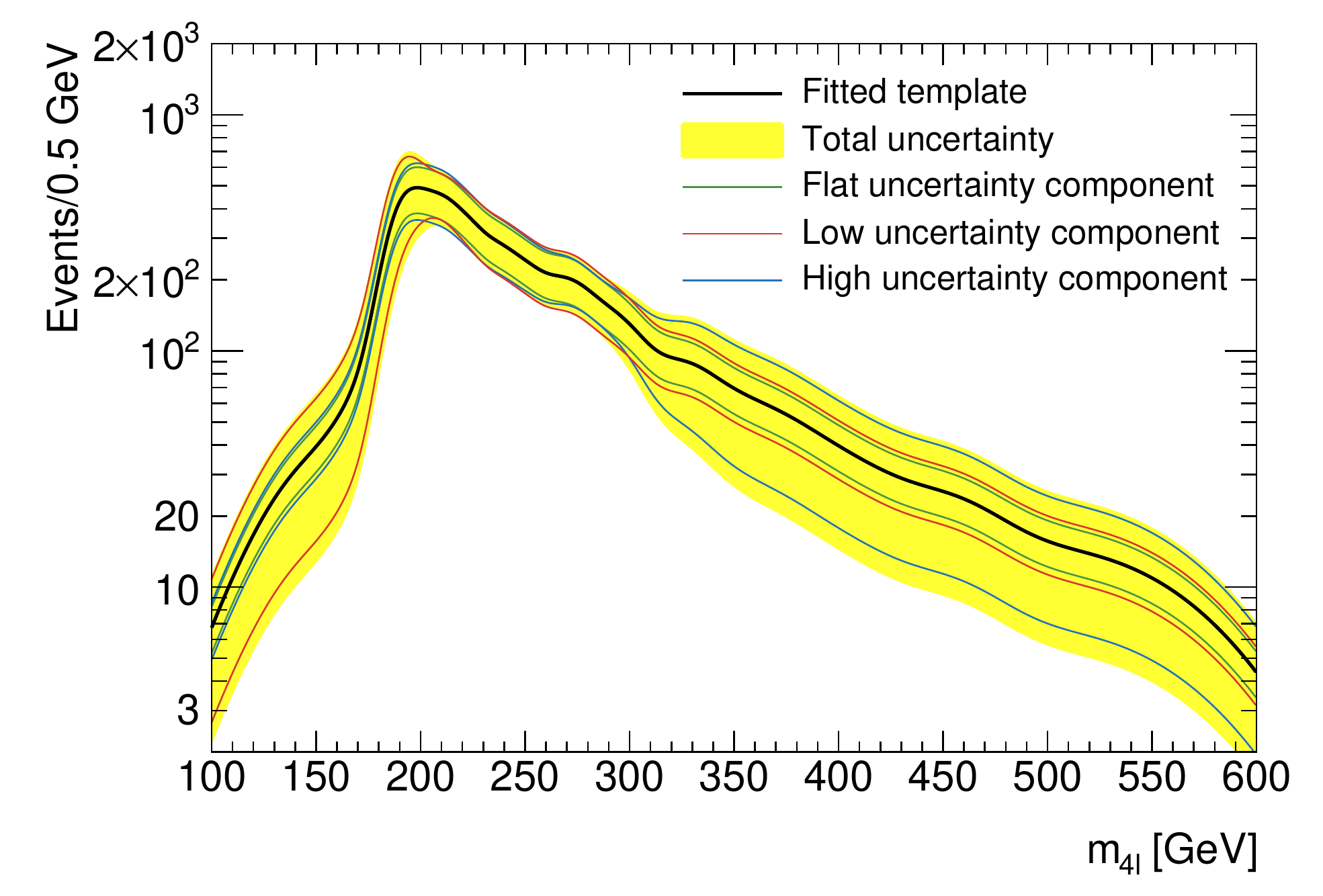}
  \end{minipage}
  \caption{Example application of the star-morphed p.d.f. to describe the effect of different uncertainties. Three example uncertainties affecting different regions of the shown spectrum have been added to the star-morphed p.d.f. The black curve shows the fitted template for the $q\bar{q}\to ZZ$ process along with the total uncertainty derived from all included sources indicated by the yellow band. The green, red and blue curves show the partial uncertainties corresponding to a flat, low mass and high mass effect, respectively.
  }
  \label{fig:starsystematics}
\end{figure}

Fig.~\ref{fig:starsystematics} shows an example application of the star-morphed p.d.f.,
used to describe the dominant background process for a typical $H \rightarrow 4l$ analysis,
labeled $q\bar{q}\to ZZ$.
As in Section~\ref{sec:oneparameter}, this background prediction has been obtained
from events simulated with MadGraph~\cite{Alwall:2014hca} and PGS~\cite{PGSWeb}.
For illustration, three non-physical systematic uncertainties have been added to the star-morphed p.d.f. One is a shape variation in the low mass region, the
second is a shape variation in the high mass region, and the third is an overall normalization uncertainty that
affects the entire mass region. The effects of these systematic uncertainties on the template distribution are shown in Fig.~\ref{fig:starsystematics}.

%% file: accuracy.tex
\section{Accuracy of moment morphing and comparison to alternative morphing algorithms }
\label{sec:accurary}

The accuracy of a moment morphing function $f_{\rm pred}(\mathbf{x}|\alpha)$ is assessed here
using a series of benchmark models for which the true distribution $f_{\rm true}(\mathbf{x}|\alpha)$ is defined.

A priori one can demonstrate that moment morphing provides an exact solution for any set of true distributions
in which the first and second moment (\emph{i.e.} the mean and width) of $f_{\rm true}(\mathbf{x}|\alpha)$ change
linearly with the morphing parameter $\alpha$ and all other moments are constant
(\emph{i.e.} the shape of the distribution does not change other than through its first and second moment).
The simplest example of such a distribution is a Gaussian distribution with a linearly changing mean and width.
However, the class of true distributions that are exactly reproduced by moment morphing is not restricted to this
example: any distribution with linearly changing first and second moments and fixed higher-order
moments is exactly described.
In particular, when considering use cases in particle physics, the physics parameters of interest
(particle masses, resonance widths, etc.) are often related to the first and second moments;
hence a correct description of these is most critical.

Deviations from the exact solution can occur in two ways: a) the dependence of the first and second moment of
$f_{\rm true}(\mathbf{x}|\alpha)$ become non-linear in $\alpha$, or b) higher moments of the shape of the distribution depend
on $\alpha$, either linearly or non-linearly.
The magnitude of both types of deviations are under explicit control of the user,
as the moment morphing algorithm allows templates to be specified at any number of freely choosable values of $\alpha$.

Deviations of the first type are generally unproblematic as it is always possible to construct a configuration of
templates in which: a) the dependence of the first and second moments between any pair of adjacent templates is
sufficiently close to linear (when choosing the piece-wise linear interpolation option),
or b) where the dependence of the moments over the entire domain of $\alpha$ is well approximated by a $n^{th}$-order
polynomial for $n$ templates (when choosing the non-linear interpolation option).
For this reason, no attempt is made here to quantify the accuracy of moment morphing due to (local)
non-linearity of the first and second moment of $f_{\rm true}(\mathbf{x}|\alpha)$.

Deviations of the second type, introduced by changing higher-order moments, may be more limiting in cases
where higher-order moments of the true distribution change rapidly as a function of $\alpha$.
It should be noted that changes in the truth distribution related to higher-order moments are empirically accounted
for in the morphing algorithm, by gradually changing the weight of moment-adjusted input templates
as a function of $\alpha$.
However, the morphed distribution does not guarantee a linear change in these higher-order moments as a function of $\alpha$,
as is done for the first and second moments.

The impact of this empirical modelling of changes in higher-order moments is quantified in the following sub-sections.

\subsection{Performance on benchmark models}

The accuracy of moment morphing is evaluated here on nine analytical benchmark models
that are similar to various particle physics use cases, and the accuracy is compared to
two other morphing algorithms that have historically been used in particle physics:
vertical morphing and ``integral morphing''.

In the vertical morphing approach\footnote{Note that the moment morphing algorithm reduces to vertical morphing strategy
if the adjustment step of the first and second moments of the input templates is omitted.} templates are interpolated
with a simple weighting strategy:
\begin{equation*}
V(x|\alpha) = (1-\alpha)/2 \cdot T_L(x) + (\alpha-1)/2 \cdot T_R(x)
\end{equation*}
where $T_{L/R}$ are the left and right template models corresponding to $\alpha = \pm 1$.
The vertical morphing approach is widely used in LHC physics analyses, notably in the modeling of distributions
in the discovery analysis of the Higgs boson.
In contrast, in the integral morphing approach~\cite{Read:1999kh} interpolation occurs between the cumulative distribution
functions of the templates $T_{L/R}$.
The integral morphing approach is more suited to models with rapidly shifting means, like moment morphing and
unlike vertical morphing, but is computationally intensive due to (numeric) integration and root-finding steps,
and is restricted to the description of one-dimensional distributions.
The integral morphing approach has been used, among others, in physics analyses published by the D0 and CDF collaborations.

\begin{table}[!t]
\begin{center}
\begin{tabular}{rlll}
Name & Model description & \multicolumn{2}{c}{Dependence on morphing parameters} \\
\hline
$N_{\mu}$ & ${\rm Gaussian}(x|\mu,\sigma=1)$ & $\mu = 2 \cdot \alpha$ \\
$N_{\sigma}$ & ${\rm Gaussian}(x|\mu=0,\sigma)$ & $ \sigma = 1 + 0.5 \cdot \alpha$ \\
$N_{\mu\sigma}$ & ${\rm Gaussian}(x|\mu,\sigma)$ & $\mu = 2 \cdot \alpha$ & $ \sigma = 1 + 0.5 \cdot \alpha$ \\
$\Gamma_{k}$ &  ${\rm GammaDist}(x|k,\theta=1)$ & $k = 3 + 0.7 \cdot \alpha$ \\
$\Gamma_{\theta}$ & ${\rm GammaDist}(x|k=3,\theta)$ & $ \theta = 1 + 0.7 \cdot \alpha$ \\
$\Gamma_{k\theta}$ & ${\rm GammaDist}(x|k,\theta)$ & $k = 3 + 0.7 \cdot \alpha$ & $ \theta = 1 + 0.7 \cdot \alpha$ \\
$C_{1}$ & ${\rm Chebychev}(x|a_1,a_2=0)$ & $a_1 = 0.5 + 0.4 \cdot \alpha$ \\
$C_{2}$ & ${\rm Chebychev}(x|a_1=0.5,a_2)$ & $ a_2 = 0.4 \cdot \alpha$ \\
$C_{12}$ & ${\rm Chebychev}(x|a_1,a_2)$ & $a_2 = 0.5 + 0.4 \cdot \alpha$ & $ a_2 = 0.4 \cdot \alpha$ \\
\hline
\end{tabular}
\end{center}
\caption{\label{tab:benchmodels}
Benchmark models used to quantify accuracy of moment morphing and alternative morphing algorithms.}
\end{table}
The nine analytical benchmark models tested here are detailed in Table~\ref{tab:benchmodels}.
For the first three benchmark models $N_{\mu},\, N_{\sigma}$ and $N_{\mu\sigma}$, based on the normal distribution, moment morphing provides an exact solution,
and these models are included in the benchmark to facilitate accuracy performance comparisons with the other morphing approaches.
The second set of benchmark models is based on the Gamma distribution:
\begin{equation*}
\Gamma(x|k,\theta) = \frac{x^{k-1}e^{-x/\theta}}{\theta^k \Gamma(k)}\,,
\end{equation*}
and are included as an example of a distribution where also higher-order moments of the true distribution changes as a function of the morphing parameter.
The last set of three benchmark models is based on the $2^{\rm nd}$-order Chebychev polynomials:
\begin{equation*}
C(x|a_1,a_2) = 1 + a_1 x + a_2 (2x^2-1)\,,
\end{equation*}
as a distribution that is representative of typical background distributions.

\begin{figure}[p]
  \centering
  \begin{minipage}[b]{1.0\linewidth}
    \begin{minipage}[b]{.49\linewidth}
      \centering
      \includegraphics[width=\textwidth]{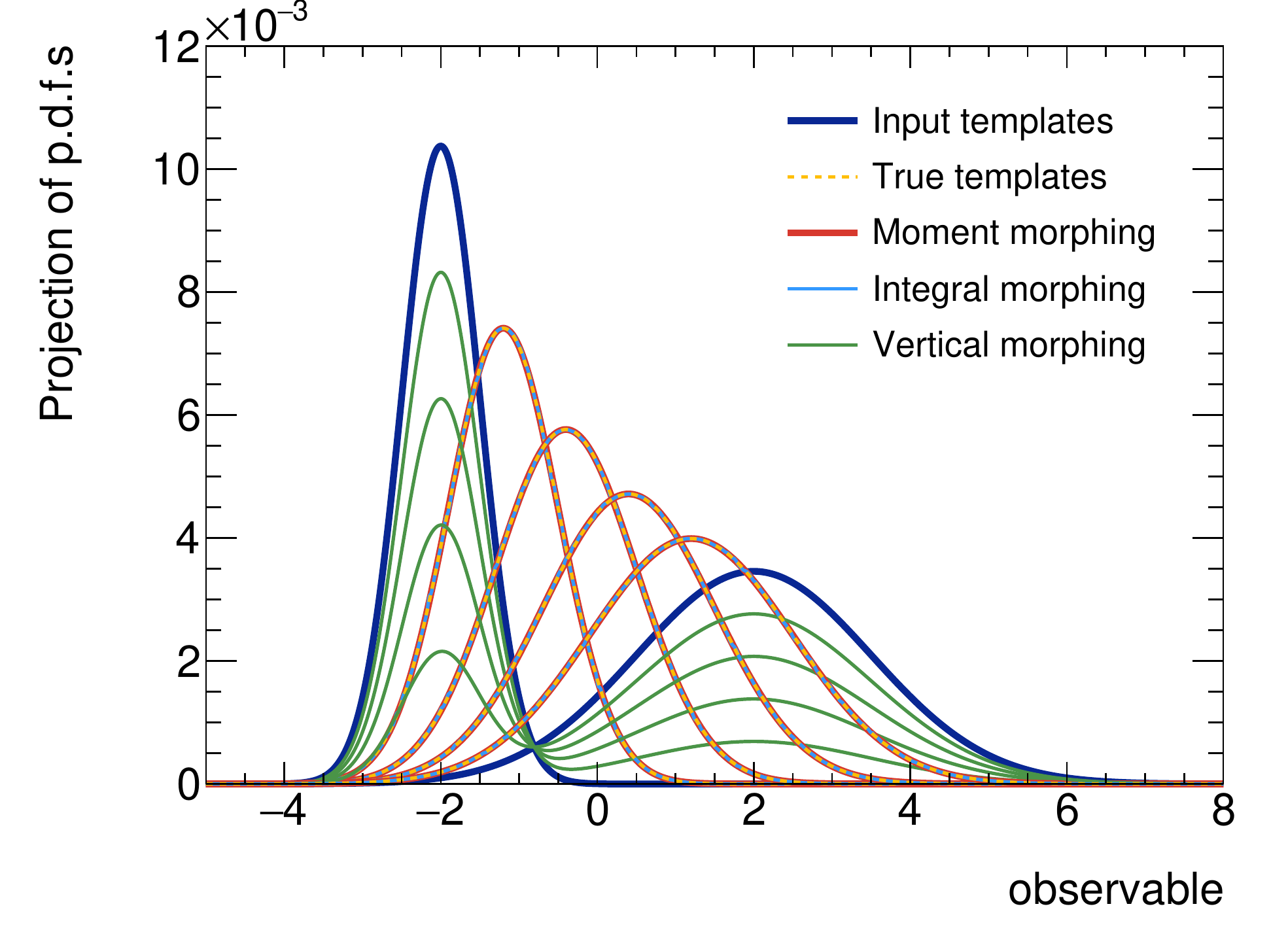}
    \end{minipage}%
    \begin{minipage}[b]{.49\linewidth}
      \centering
      \includegraphics[width=\textwidth]{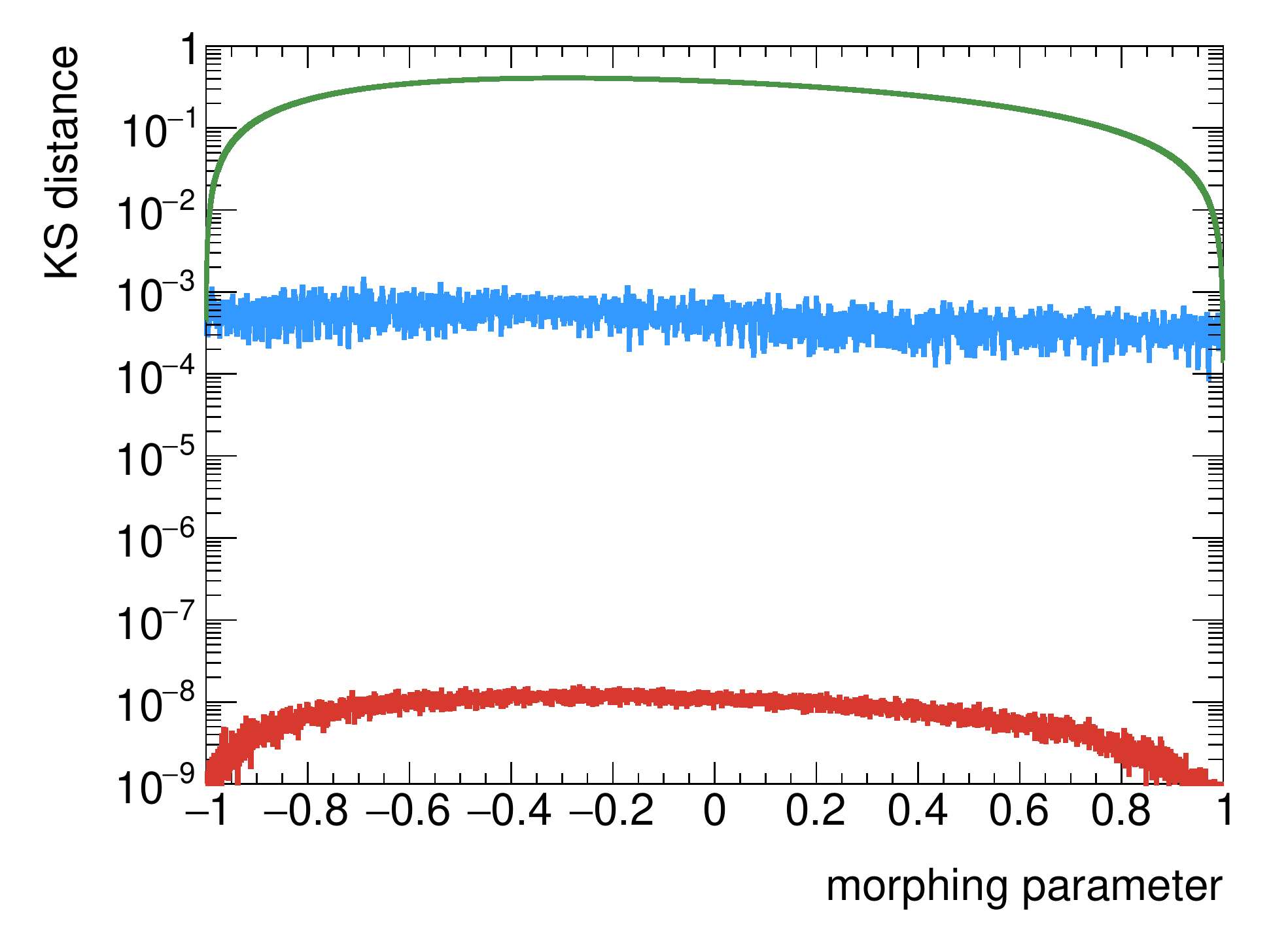}
    \end{minipage}
  \end{minipage}
  \begin{minipage}[b]{1.0\linewidth}
    \begin{minipage}[b]{.49\linewidth}
      \centering
      \includegraphics[width=\textwidth]{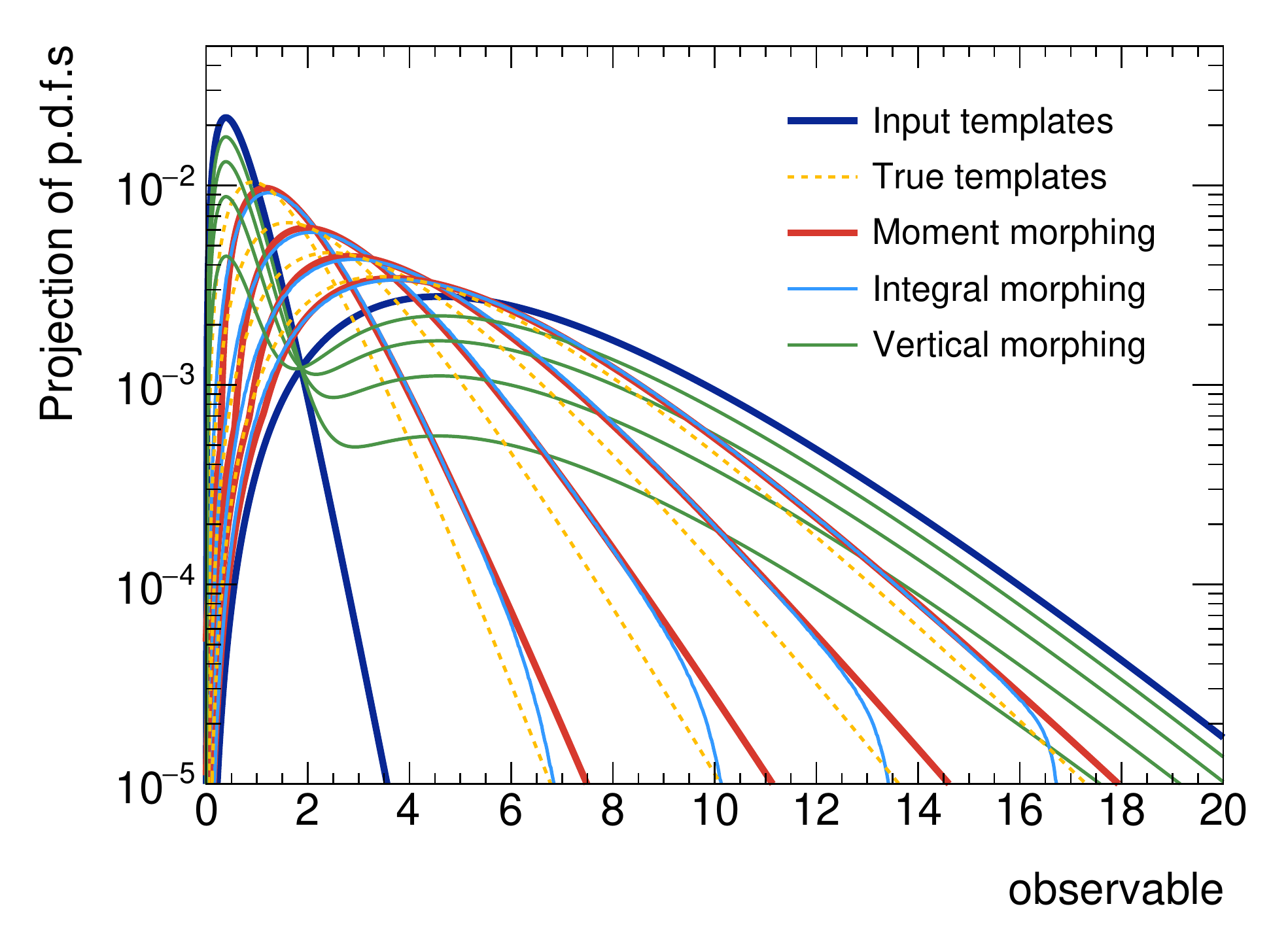}
    \end{minipage}%
    \begin{minipage}[b]{.49\linewidth}
      \centering
      \includegraphics[width=\textwidth]{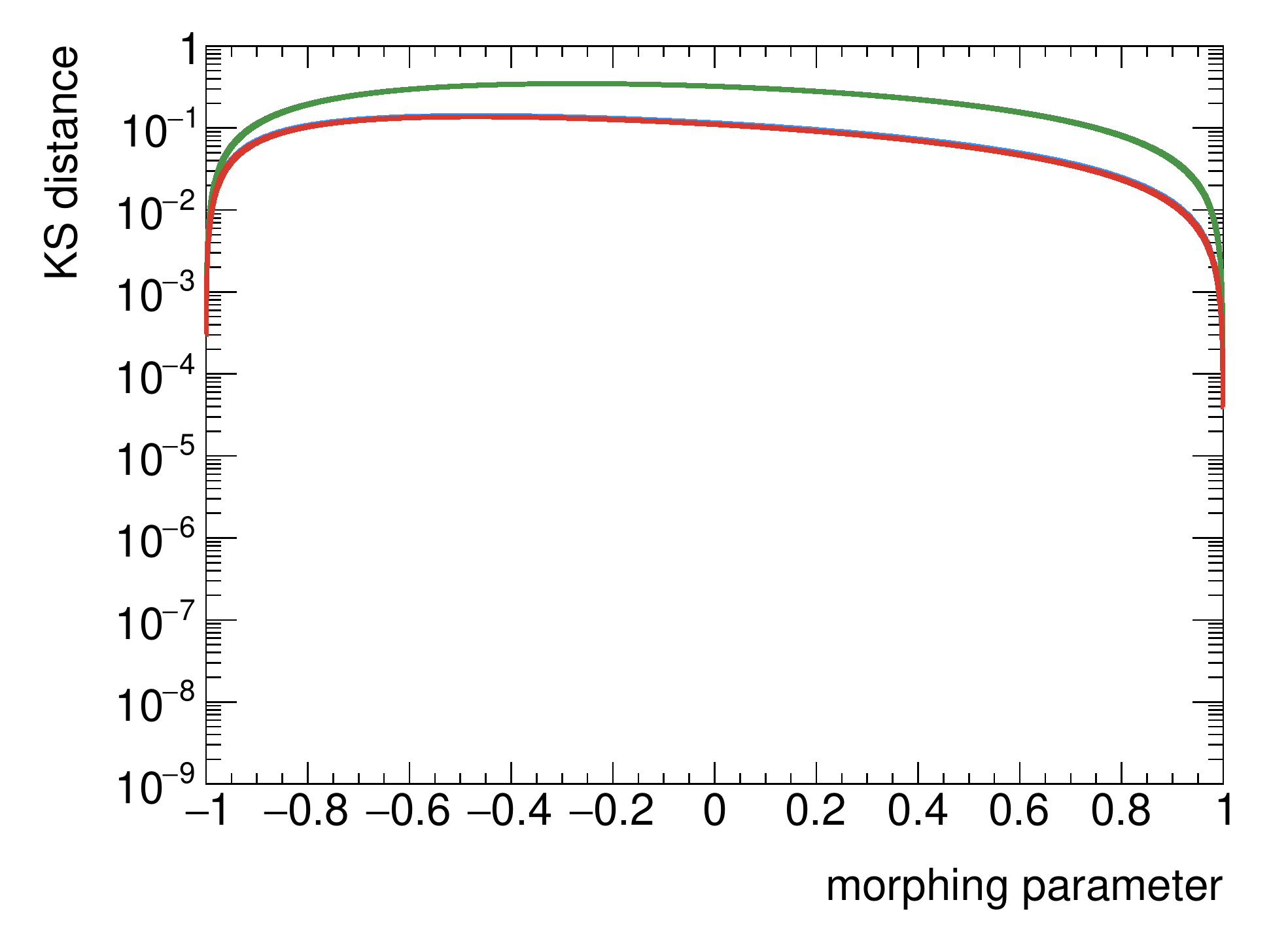}
    \end{minipage}
  \end{minipage}
  \begin{minipage}[b]{1.0\linewidth}
    \begin{minipage}[b]{.49\linewidth}
      \centering
      \includegraphics[width=\textwidth]{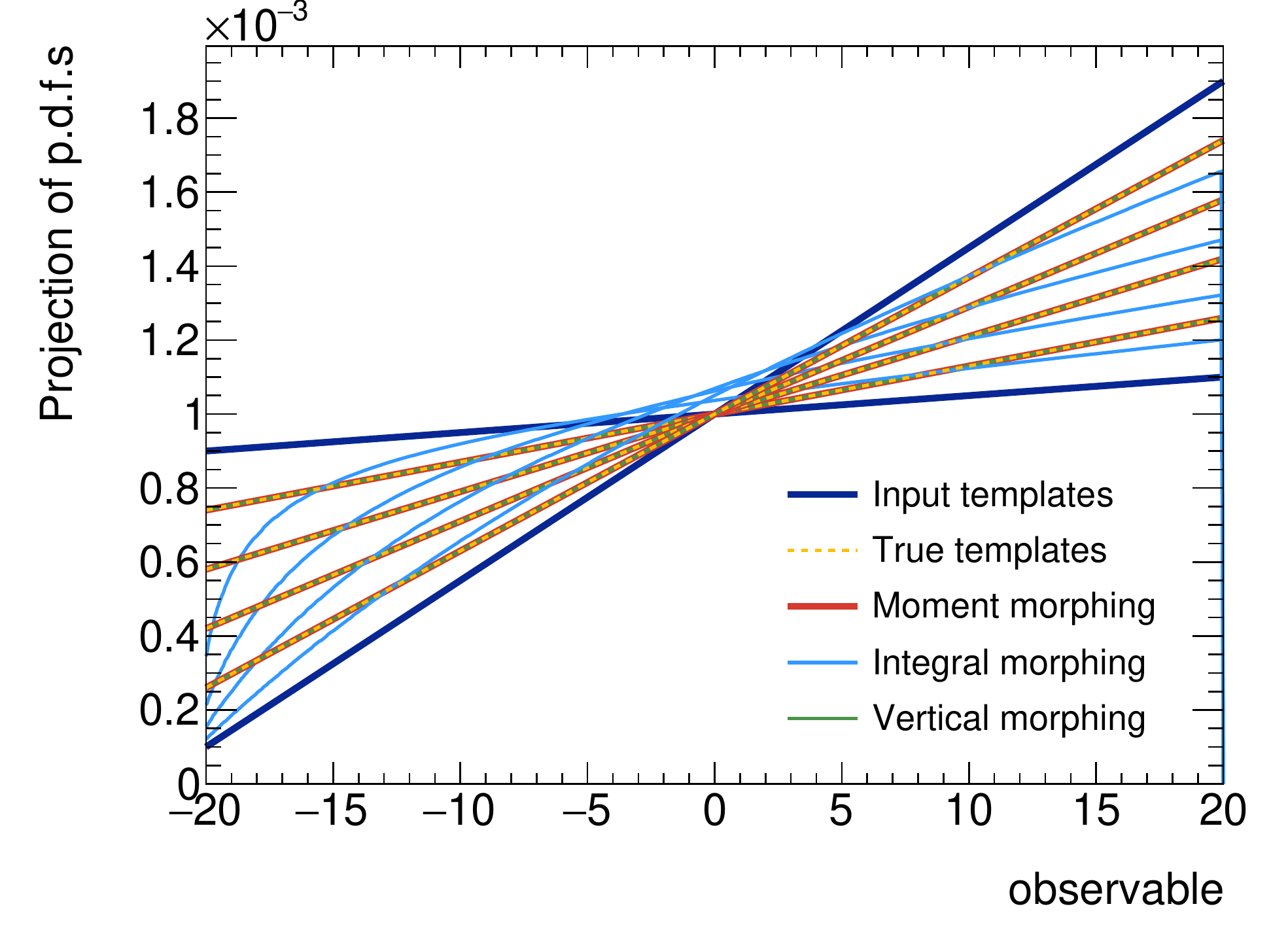}
    \end{minipage}%
    \begin{minipage}[b]{.49\linewidth}
      \centering
      \includegraphics[width=\textwidth]{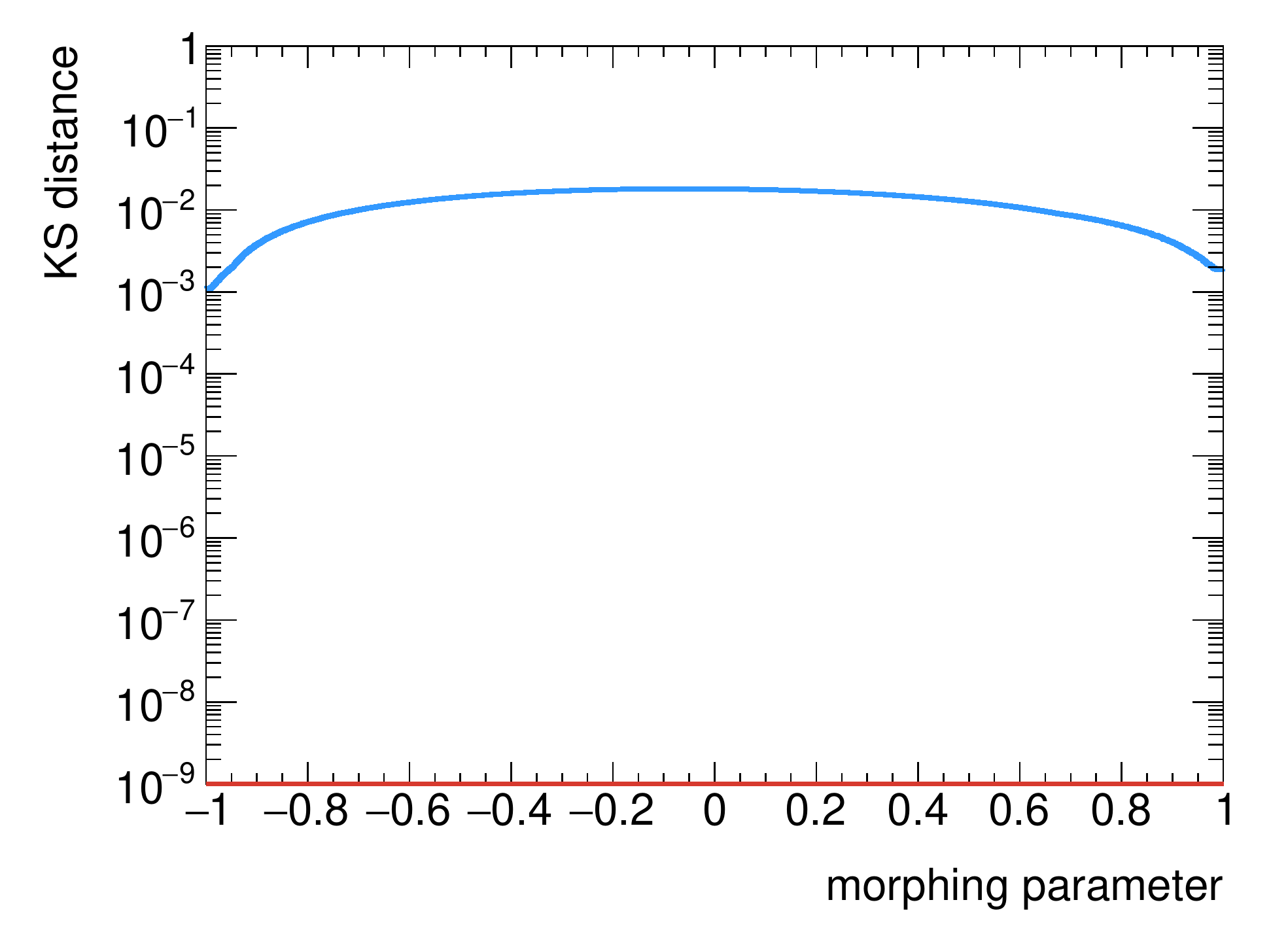}
    \end{minipage}
  \end{minipage}
  \caption{\label{fig:accu_bench}
Accuracy of the moment morphing, vertical morphing and integral morphing algorithms on three of the benchmark models defined
in Table~\ref{tab:benchmodels}. Top row: the $N_{\mu\sigma}$ model, center row: the $\Gamma_k$ model, bottom row: the $C_1$ model.
The plots in the left column overlay the truth model distributions for various values of the morphing parameter $\alpha$ ($\alpha=\pm1$ shown in dark blue, which also serve as templates for the morphing algorithms, other values of $\alpha$ shown in yellow dashed) on the predicted distribution by moment morphing (red), vertical morphing (green) and integral morphing (light blue).
To quantify the level of agreement between truth and morphing predictions, the right column shows the KS distance between the truth model and predicted distributions
as a function of the morphing parameter $\alpha$ by the three types of morphing models, using the same colour codings as the left column plots.
Some of the morphing predictions overlap (almost) perfectly, making them difficult to distinguish visually. They are discussed further in the text.}
\end{figure}

Given a pair of input templates positioned at $\alpha = \pm 1$, the accuracy of the morphing methods is quantified using the Kolmogorov-Smirnov (KS) distance.
This is the largest distance between the cumulative distribution functions of the truth model and the cumulative distribution function of the morphing interpolation model
for any value of $x$, as a function of the morphing parameter $\alpha$.
The KS distance values range between zero (perfect agreement between the two distributions) and one (complete spatial separation of the distributions in $x$).

\begin{table}[!t]
\begin{center}
\begin{tabular}{r|cc|cc|cc|}
           & \multicolumn{2}{c|}{Moment morphing} & \multicolumn{2}{c|}{Vertical morphing} & \multicolumn{2}{c|}{Integral morphing} \\
Name & $KS(\alpha=0)$ & $KS_{max}$ & $KS(\alpha=0)$ & $KS_{max}$ & $KS(\alpha=0)$ & $KS_{max}$ \\
\hline
$N_{\mu}$                 & $0$ & $0$ & $0.28$ & $0.28$ & $4.8 \cdot 10^{-4}$ & $8.1 \cdot 10^{-4}$ \\
$N_{\sigma}$            & $3.7 \cdot 10^{-9}$ & $5.3 \cdot 10^{-9}$ & $0.044$ & $0.046$ & $2.8 \cdot 10^{-4}$ & $9.1 \cdot 10^{-4}$ \\
$N_{\mu\sigma}$       & $1.1 \cdot 10^{-8}$ & $1.3 \cdot 10^{-8}$ & $0.36$ & $0.40$ & $4.4 \cdot 10^{-4}$ & $9.2 \cdot 10^{-4}$ \\
\hline
$\Gamma_{k}$           & $0.0064$ & $0.0069$ & $0.032$ & $0.032$ & $0.0033$ & $0.0036$ \\
$\Gamma_{\theta}$   & $9.2 \cdot 10^{-4}$ & $1.1 \cdot 10^{-3}$ & $0.28$ & $0.32$ & $5.4 \cdot 10^{-4}$ & $6.9 \cdot 10^{-4}$ \\
$\Gamma_{k\theta}$ & $0.11$ & $0.14$ & $0.32$ & $0.35$ & $0.11$ & $0.14$  \\
\hline
$C_{1}$                     & $0$ & $0$ & $0$ & $0$ & $0.018$ & 0.018 \\
$C_{2}$                     & $0.0086$ & $0.0086$ & $0.0086$ & $0.0086$ & $0.013$ & $0.013$ \\
$C_{12}$                   & $0.020$ & $0.020$ & $0.020$ & $0.020$ & $0.040$ & $0.040$ \\
\hline
\end{tabular}
\end{center}
\caption{\label{tab:benchresults}
Accuracy of the moment morphing, vertical morphing and integral morphing algorithms on the nine benchmark models define
in Table \ref{tab:benchmodels}, expressed in the KS distance at $\alpha=0$ and the largest KS distance that occurs in the range
$-1<\alpha<1$.}
\end{table}
Fig.~\ref{fig:accu_bench} illustrates the accuracy of the three morphing approaches on the $N_{\mu\sigma}, \Gamma_{k\theta}$ and $C_{1}$ models,
along with the KS statistic as a function of $\alpha$ for all morphing approaches.
Accuracy metrics for all nine benchmark models are given in Table \ref{tab:benchresults}, which summarizes for each benchmark model
and each morphing strategy the KS distance at $\alpha=0$ (the morphing mid-point), and the worst KS distance that occurs in the morphing range $\alpha \in [-1,1]$.

As expected, the moment morphing approach exactly replicates the $N_{\mu\sigma}$ benchmark model, unlike the vertical morphing approach.
The integral morphing algorithm can also exactly replicate the $N_{\mu\sigma}$ model, but is limited to an accuracy of about 0.001 due to
numerical precision limitations, related to integration and root-finding in the algorithm's implementation.
For the $\Gamma_{k\theta}$ benchmark model, no morphing algorithm can exactly reproduce the truth model due to changing higher-order moments,
but the moment and integral morphing algorithms perform about equally, and significantly better than the vertical morphing algorithm.
For the $C_1$ benchmark model both the moment morphing and vertical morphing perform identically, as the first and second moment of the
true distribution do not depend on $\alpha$, hence both approaches reduce to the same algorithm, which happens to describe the $C_1$ model
perfectly\footnote{This perfect performance is specific to the $C_1$ benchmark that changes the first Chebychev coefficient only.
Truth models that change high-order coefficients are imperfectly described my all morphing algorithms, as detailed in Table~\ref{tab:benchresults}.}.

In summary, in these tests the moment morphing approach is as good as and usually better than the two alternative morphing methods.

\subsection{Accuracy for multi-dimensional distributions and multi-parameter morphing} \label{sec:accumult}

For morphing models with multiple morphing parameters, and/or with input distributions with multiple observables,
it is challenging to capture the accuracy with a limited set of benchmark models 
as KS tests have no simple equivalent for multidimensional distributions, 
and the number of benchmark permutations to test becomes large with multiple morphing parameters.

Nevertheless, one can illustrate the challenges of morphing and the performance of moment morphing with a single simple yet
challenging 2-parameter benchmark model for 2-dimensional distributions: a multivariate normal distribution:
\begin{equation*}
f_{\rm true}(\mathbf{x}|\alpha,\beta)={\rm Gaussian}(\mathbf{x}|\vec{\mu},{\bf V})\,,
\end{equation*}
for which all five degrees of freedom, $\mu_x, \mu_y, \sigma_x \equiv V_{xx}, \sigma_y \equiv V_{yy}$
and $\rho \equiv V_{xy}/\sqrt{V_{xx} V_{yy}}$ depend linearly on two morphing parameters $\alpha,\beta$.

In the case where the correlation coefficient $\rho$ is independent of $\alpha,\beta$, moment morphing provides
the exact solution for this multi-variate Gaussian model, similar to the one-dimensional case, even if the dependence
of $\mu_x, \mu_y, \sigma_x, \sigma_y$ is strong, as long as it is linear.
Conversely, vertical morphing will perform poorly, especially if $\mu_x$ or $\mu_y$ depend strongly on $\alpha,\beta$,
while integral morphing is not available for multi-dimensional distributions.

The more challenge scenario where also $\rho$ depends on $\alpha,\beta$ is visualized in Fig.~\ref{fig:example2}.
Moment morphing will not provide an exact solution for this class of models as the covariance moments are not explicitly corrected for,
nevertheless the covariance of the interpolated shape at $(\alpha,\beta)=(0.5,0.5)$ reasonably matches the covariance of the true model,
although the shape is no longer perfectly Gaussian.

%% file: implementation.tex
\section{Implementation}
\label{sec:implementation}

This section discusses the publicly available morph classes, and is followed by details of
the chosen extrapolation approach of a morph parameter beyond the provided input range.

\subsection{Available morph classes}
\label{sec:classes}

Moment morphing has been implemented in \CC~for the \texttt{RooFit} toolkit~\cite{Verkerke:2003ir}.
As of \texttt{ROOT} release 5.34.22, the morphing features described in this document are available in the \texttt{RooFit} models library.
Common to all classes is the ability to handle one or multiple observables,
as well as the implementation of a cache that stores pre-calculated expensive components such as numerically computed moments,
\emph{e.g.} the means and widths of each input template, required for the translation of the corresponding observables.

The following moment morph classes are available in \texttt{RooFit}.
\begin{itemize}
\item The \texttt{RooMomentMorph} p.d.f. can be used to interpolate between an arbitrary number of reference distributions
using a single morphing parameter.
The algorithms settings described in this paper, \emph{e.g.} linear 
or truly non-linear, can be used.

Furthermore, a sine-linear variant transforms $ m_\text{frac}$ to $\sin (\pi/2\cdot m_\text{frac})$ before calculating the coefficients, thus ensuring a continuous and differentiable transition when crossing between two adjacent sets of enclosing grid points.
In addition, non-linear coefficients for adjusting the moments of the p.d.f.s can be mixed with linear coefficients
when constructing the morph p.d.f., and an option is available to select positive non-linear coefficients only.
\item \texttt{RooStarMomentMorph} is the natural extension for combining multiple one-dimensional RooMomentMorph p.d.f.s with one common sampling point.
The class supports linear and sine-linear interpolation.
The transformation of the template observables can be turned on or off.
\item \texttt{RooMomentMorphND} allows the parametrization of a $n$-dimensional parameter space,
interpolating linearly or sine-linearly between reference points sitting on a hyper-cube or arbitrary size.
\item \texttt{Roo1DMomentMorphFunction} and \texttt{Roo2DMomentMorphFunction} are similar to the top moment morph p.d.f.,
but can be used to interpolate between functions, not p.d.f.s. Available for one or two morph parameters.
\end{itemize}

Example code of how to use the moment morph p.d.f. is given below. 
\\
\begin{lstlisting}[caption=Sample code to build the model shown in Fig.~\ref{fig:construction}.] %\label{list:examplecode}

// This example builds two normal distributions and uses moment morphing
// to interpolate between the templates using RooFit.
using namespace RooFit;

// Create a persistable container for RooFit projects, allowing to use 
// a simplified scripting language to build the p.d.f.s needed in this 
// example.
RooWorkspace w("w", 1);

// Build two normal distributions, corresponding to different values in 
// the morph parameter space. They share the same observable, but have 
// otherwise different moments, i.e. mean and width.
w.factory("RooGaussian::gaussian1(obs[0,400],50,15)");
w.factory("RooGaussian::gaussian2(obs,300,40)");

// Build a RooMomentMorph p.d.f. which interpolates between the normal 
// distributions created before. The interpolation is parametrized by 
// the parameter alpha and the reference templates map to alpha=0 and 
// alpha=1 respectively.
w.factory("RooMomentMorph::morphpdf(alpha[0,1],obs,
                                    {gaussian1,gaussian2},{0,1})");

// Set the morphing parameter alpha explicitly to 0.5.
w::alpha->setVal(0.5);

// Create a frame to draw the p.d.f.s from before and show the input 
// templates as solid blue curves and the moment morph pdf at alpha=0.5 
// in dashed red.
RooPlot* frame = w::obs->frame();
w::gaussian1->plotOn(frame, LineColor(kBlue), LineStyle(kSolid));
w::gaussian2->plotOn(frame, LineColor(kBlue), LineStyle(kSolid));
w::morphpdf->plotOn(frame, LineColor(kRed), LineStyle(kDashed));
frame->Draw();
\end{lstlisting}

\input extrapolation

\subsection{Computational performance of moment morphing}

\begin{figure}[!t]
  \centering
  \begin{minipage}[b]{.6\linewidth}
    \centering
    \includegraphics[width=\textwidth]{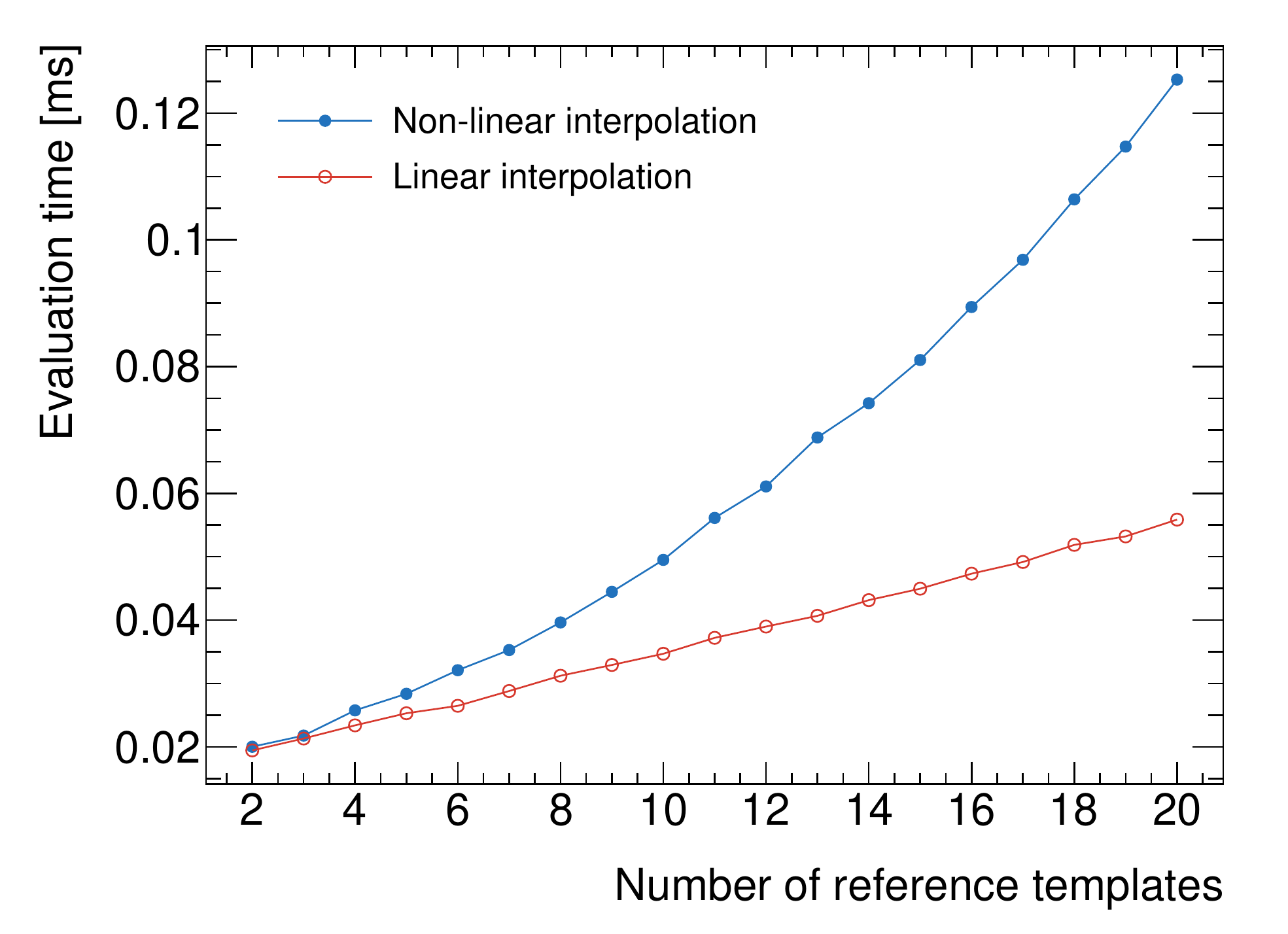}
  \end{minipage}%
  \caption{Benchmark results for \texttt{RooMomentMorph}. The figure shows the average CPU time for the evaluation of the morph p.d.f. as a function of the reference templates used for the interpolation for linear (red) and non-linear (blue) algorithms settings. This was measured on an Intel(R) Xeon(R) CPU E5-2450 @ $\unit[2.10]{GHz}$ with $\unit[2]{GB}$ of memory per core. As reference p.d.f.s normal distributions with varying mean are used.}
  \label{fig:benchmark}
\end{figure}

Template morphing is one of the computationally limiting factors in current HEP analyses, which makes understanding the performance of the algorithm at hand crucial. Fig.~\ref{fig:benchmark} compares the performance of the linear and non-linear algorithms settings in terms of average CPU time needed for the evaluation of the morph p.d.f. as a function of the used reference templates. The benchmark makes use of the caching described in section~\ref{sec:classes}. It
excludes the computation time needed for the calculation of the moment integrals over the input templates,
which is a one-time calculation, the result of which can be cached in the RooFit workspace file along with the model
if desired by the user.
Details of the setup are described in the caption of the figure.

For the interpretation of performance numbers it should be noted that the evaluation time for
non-linear
morphing models depends quadratically on the number of reference templates $n$,
as for every template the associated coefficient is a product of $n$ distances.
In the linear case, the evaluation time is driven by the efficiency of the algorithm finding the two
closest reference points surrounding the point at which the p.d.f. is evaluated.
For $10$ ($20$) reference templates, the ratio of non-linear over linear evaluation time is $1.4$ ($2.2$).

%% file: extrapolation.tex
\subsection{Extrapolation beyond input boundaries}

By construction, the validity of the morph p.d.f. of Eqn.~\ref{eq:morphpdf}
is highest when interpolating its morph parameter(s)
within the provided range(s) of input values.
Beyond these the predictive power is {\it a priori} unknown,
but of course can be interesting to investigate.
Some {\it ad hoc} choices need to be made for extrapolation cases, 
which were not covered by the description of the algorithm so far.

As apparent from Eqn.~\ref{eq:coefficient},
when extrapolating $m$ beyond the input boundary values $m_\text{min}$ and $m_\text{max}$,
the coefficients $c_i(m)$ increase in size and may become highly negative.
In this situation the morph p.d.f. can become smaller than zero, and as such ill-defined.
To prevent this, the following extrapolation approach is implemented.
\begin{itemize}
\item Whenever one of the morph parameters extends beyond the input range,
\emph{i.e.} $m\!<\!m_\text{min}$ or $m>m_\text{max}$, the coefficient multiplied with
the nearest input p.d.f. is forced to one, and all other coefficients are set to zero.
\item The same is done for the transformed width of Eqn.~\ref{newmean},
which is to remain greater than zero.
\item Beyond the input boundaries, the transformed mean of Eqn.~\ref{newmean} does remain well-defined,
and uses the coefficients of either Eqn.~\ref{eq:coefficient} or Eqn.~\ref{eq:coefficientlinear},
depending on the linearity setting used.
\end{itemize}

%% file: conclusion.tex
\section{Conclusion}

We have described a new algorithm to interpolate between Monte Carlo sample
distributions, called moment morphing.  
The proposed technique is based on a linear combination of the input templates, 
has a non-linear dependence on the parameters that control the morphing interpolation,
and can be constructed in any orthogonal basis.

Compared with existing methods, it allows for both binned histogram and continuous templates,
has both horizontal and vertical morphing,
and is not restricted in the number of input templates, observables, or model parameters.
In particular, the latter feature allows the moment morphing technique to model the impact of
a non-factorizable response between different model parameters.
The moment-morphed p.d.f. is self-normalized, and therefore, in terms of speed and stability, 
scales well with the number of input templates used.
This is a useful feature in modern day particle physics where large numbers of templates are often required to model
the variations by systematic uncertainties.

In a suite of comparison tests the moment morphing method proofs to be as accurate as, or better than,
two commonly used alternative techniques.

Various implementations of the moment morphing technique are publicly available through the \texttt{RooFit} toolkit.